\documentclass[12pt]{article}
\usepackage[utf8x]{inputenc}
\usepackage[T1]{fontenc}
\usepackage{amssymb,amsmath,mathtools,amsfonts}
\usepackage{graphicx}
\usepackage{lmodern}
\usepackage{float}
\usepackage[colorlinks=true, allcolors=blue]{hyperref}
\usepackage{tabularx,booktabs}
\usepackage{abstract}
\usepackage{caption}
\usepackage{titlesec}

\usepackage[numbers,sort&compress]{natbib}
\evensidemargin\oddsidemargin \hoffset-22.4mm \voffset-28.4mm
\title{Register jumps on the clarinet: numerical and in-vitro investigation into  basins of attraction and phase-tipping}
\author{Nathan Szwarcberg\textsuperscript{a, b}\footnote{Corresponding author. \\
E-mail address: nathan.szwarcberg@buffetcrampon.com (N.\ Szwarcberg)}, 
Tom Colinot\textsuperscript{a}, 
Christophe Vergez\textsuperscript{b}, 
Michaël Jousserand\textsuperscript{a}\\
Léonie Maignan\textsuperscript{b}, 
Anthia Patsinakidou\textsuperscript{b},\\
Giordano Gatti\textsuperscript{b},
Hrant Arzumanyan\textsuperscript{b},
Pedro Faria Oliveira Morais\textsuperscript{b}\\
\small
\textsuperscript{a} Buffet Crampon, 5 Rue Maurice Berteaux, 78711 Mantes-la-Ville, France\\
\small
\textsuperscript{b} Aix Marseille Univ, CNRS, Centrale Med, LMA, Marseille, France\\
}
\date{\today}

\begin{document}
\maketitle
\begin{onecolabstract}
When playing the clarinet, opening the register hole allows for a transition from the first to the second register, producing a twelfth interval. On an artificial player system, the blowing pressure range where the second register remains stable can be determined by gradually varying the blowing pressure while keeping the register hole open. However, when the register hole is opened while the instrument is already producing the first register, the range of blowing pressures that lead to a stable second register is narrower than the full stability zone of the second register.
This phenomenon is investigated numerically by performing multiple hole openings at different times, for various values of the blowing pressure and the embouchure parameter. In some narrow regions of the control parameter space, the success of a register transition depends on the phase at which the hole is opened. This illustrates an instance of phase-tipping, where the limit cycle of the closed-hole regime may intersect multiple basins of attraction associated with the open-hole regimes. Furthermore, to assess the robustness of the basins of attraction, random noise is introduced to the control parameters before the register hole is opened. Results indicate that the equilibrium regime is more robust to noise than the other oscillating regimes. Finally, long-lasting transient quasiperiodics are investigated. The phase at which the hole is opened influences both the transient duration and the resulting stable regime.
\end{onecolabstract}
  \vspace{10pt}

\section{Introduction}\label{sec:introduction}

When characterizing a clarinet fingering, one of the first steps consists in measuring the minimum and maximum blowing pressures that allow a note to be played, for a fixed embouchure. 
These limits are known as the oscillation and extinction thresholds \cite{dalmont2005analytical,dalmont_oscillation_2007,atig2004saturation}.
The Artificial Player System \cite{backus1961vibrations, mcginnis1943experimental, li2016effect, chatziioannou2019investigating} is commonly used to determine these limits by gradually increasing the blowing pressure.
Above the extinction threshold, the reed is pressed against the mouthpiece and stops vibrating. 
When the pressure is then reduced, the reed starts oscillating again at a lower pressure, sometimes called the ``inverse threshold'' \cite{dalmont_oscillation_2007}. 
The difference between the extinction and inverse thresholds creates a hysteresis loop, where the equilibrium (no sound) and the oscillating regimes coexist and are multistable \cite{colinot2021multistability,colinot2025cartography}.

In systems where several regimes are multistable, the regime that is actually produced depends on its basin of attraction.
The basin of attraction of a regime (a fixed point or a limit cycle) is defined as the set of initial conditions that lead the system to converge toward this regime \cite{strogatz2024nonlinear}.
In musical instruments, identifying these basins helps predict which regime a musician is most likely to play \cite{benade1996physics, colinot2021multistability, pegeot2024playability, passa2025basins, terrien2025basins}. 
However, mapping the full basin of attraction is computationally demanding because of the high dimensionality of the phase space \cite{colinot2021multistability}.
Moreover, only the regions of this phase space that correspond to actual musician gestures are of practical interest.
For the clarinet, such player-informed initial conditions remain difficult to estimate, although experimental data on attack transients have been collected \cite{li2016effect, almeida2017mechanism, pamies2020influence}.

The present study addresses these challenges by focusing on transitions between two notes.
In this context, all initial conditions for the second note belong to the limit cycle of the first note. 
We consider the particular case of opening the register hole, which induces a transition from the first to the second register, corresponding to an ascending musical twelfth.
This scenario is relevant for clarinetists, for whom mastering register changes is crucial to avoid the production of unwanted notes and to improve the smoothness of the \textit{legato} \cite{blazich2005amand}.
In this context, we investigate the multistable regions associated with the second-register regime. 
While existing models describe the effect of gradual hole openings \cite{guillemain2005dynamic,terroir2005simple,terroir2006,taillard2018phd,pamies2019player}, we here mostly study the system’s response to an instantaneous opening of the register hole. 
This framework allows us to explore \textit{phase tipping} phenomena, which occur ``when a rapid change in an input causes the system to tip into a different state, but only from certain phases'' \cite{alkhayuon2021phase}.

In a closely related context, Terrien \textit{et al.}~(2025) \cite{terrien2025basins} studied how the basin of attraction of a periodic regime at a given blowing pressure $\gamma_f$ evolves with the rate of increase of the transient starting from a lower blowing pressure $\gamma_i$. 
They highlighted a phenomenon of \textit{rate-induced tipping} \cite[p.~6]{terrien2025basins}, where the regime reached depends on the rate of change between $\gamma_i$ and $\gamma_f$.
In the present work, we also briefly investigate rate-induced tipping in the case of a progressively opened register hole.

Our investigation builds upon previous studies on the influence of localized nonlinear dissipation on the dynamic behavior of clarinet models \cite{szwarcberg2023amplitude, szwarcberg2025oscillation}, and more specifically on the register hole \cite{szwarcberg2024second, colinot2025cartography, szwarcberg2025geometric}.
In particular, in \cite{szwarcberg2024second} we investigate the ability of a register hole to trigger a transition from the first to the second register when it is opened during playing. 
In this study, musicians play a steady first-register note under constant control parameters, while the register hole is suddenly opened at random times. 
The resulting regime is then analyzed for various hole diameters and positions, both experimentally and numerically using a modal decomposition model. 
The results show that accounting for nonlinear losses make the first register lose its stability, provided the resonator geometry is accurately modeled.

The present article extends this framework but differs from \cite{szwarcberg2024second} in both experimental setup and numerical modeling. 
Experimentally, we replace musician tests with measurements on an Artificial Player System, ensuring the control of the blowing pressure and embouchure.
Numerically, we introduce a waveguide clarinet model that includes localized nonlinear losses at the register hole through a quasi-steady flow boundary condition \cite{szwarcberg2025oscillation}. 
This sparse physical model allows for a detailed exploration of the control parameter space, including the influence of the limit-cycle phase at the moment of the opening of the hole.
Moreover, it provides a practical framework to simulate continuous hole openings.

The paper is organized as follows.
Section~\ref{sec:exp} presents the experiment conducted on a cylindrical clarinet equipped with a register hole and measured using the Artificial Player System.
Blowing pressure ramps are first carried out with the register hole either closed or open, to identify stable and multistable regions. 
Then, multiple hole openings are performed to determine the blowing pressure range producing stable register jumps. 
Section~\ref{sec:model} introduces the waveguide clarinet model accounting for localized nonlinear losses in the register hole. 
Time-domain simulations analogous to the experiment are presented in Sections~\ref{sec:sim:ramps} and \ref{sec:sim:holeopenings}.
Experimental and numerical multistability regions are compared in Sections \ref{sec:results:exp} and \ref{sec:results:sim}, with particular attention to narrow regions of the control parameter space (called ``transition regions''), where the first-register limit cycle interacts with the basins of attraction of competing regimes (Section~\ref{sec:results:multi}).
Basin stability \cite{menck2013basin} is analyzed by introducing white uniform noise in the control parameters before opening the register hole (Section~\ref{sec:results:pt}), and long transient regimes are studied in Section~\ref{sec:results:transTime}. 
Finally, Section~\ref{sec:discussion} discusses the relevance of the comparison between experimental observations and model predictions, and examines the impact of continuous hole opening on the transition dynamics.

\section{Experiment}\label{sec:exp}
\begin{figure}[H]
	\centering
	\includegraphics[width=.8\textwidth]{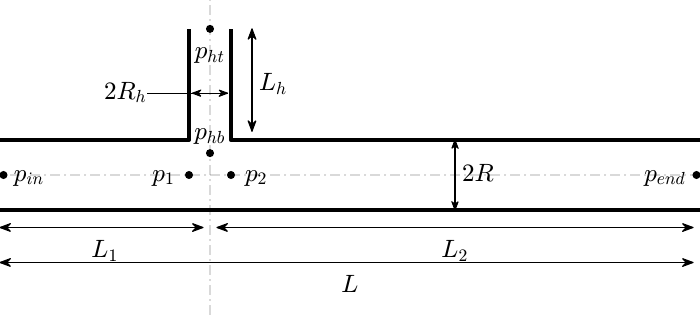}
	\caption{Definition of the digital resonator studied.}
	\label{fig:2}
\end{figure}

A simplified clarinet is built, made of a cylindrical tube of total length $L = 298$~mm and inner radius $R=7.5$~mm. 
A register tube is placed at a distance $L_1 = 132$~mm from the end of the mouthpiece. This register tube has a chimney height $L_h = 10$~mm and a diameter $d_h=2R_h = 3$~mm.
A schematic representation is shown in Figure~\ref{fig:2}.

In normal playing conditions, a clarinetist plays an E$\flat$4\footnote{All notes in this article are expressed in B$\flat$, as they would be written on a clarinet chart.} (first register, R1$^{(c)}$) when the hole is closed, and a B$\flat$5 (second register, R2$^{(o)}$) when it is open.
The superscripts $^{(o)}$ and $^{(c)}$ refer to regimes related to the open hole or the closed hole respectively\footnote{Note that it is possible to play the second register while the hole is closed (R2$^{(c)}$), and the first register when it is open (R1$^{(o)}$).}.

Measurements are performed using an artificial player system, which consists in a sealed chamber enclosing the instrument's mouthpiece and reed. 
The static pressure in the chamber, noted $P_\mathrm{blow}$, is modified through an air supply.
The reed is damped by an artificial lip, which can be positioned horizontally and vertically.

Compared to a human player, an artificial player system enables to control independently the height of reed at rest and the blowing pressure. 
Throughout the experiment, the position of the lip remains fixed.
The blowing pressure $P_\mathrm{blow}$ is controlled in two ways.
First, $P_\mathrm{blow}$ is measured through a pressure sensor (Kulite Semiconductors XT-190M-350mBARG), connected sequentially to a signal conditioner (Kulite Semiconductors KSC-2), an acquisition card (National Instruments PCIe-6341, with a resolution of 0.305~mV), and the computer.
Secondly, $P_\mathrm{blow}$ is manually modified by a pressure reducer (RS PRO 11-818), directly connected to the pressurized air supply.

Finally, an external microphone measures the acoustic field radiated by the instrument, noted $P_\mathrm{out}$.
Measurements are performed at a sampling frequency of 32~kHz.

\subsection{Pressure ramps}

\begin{figure}[H]
	\centering
	\includegraphics[width=.6\textwidth]{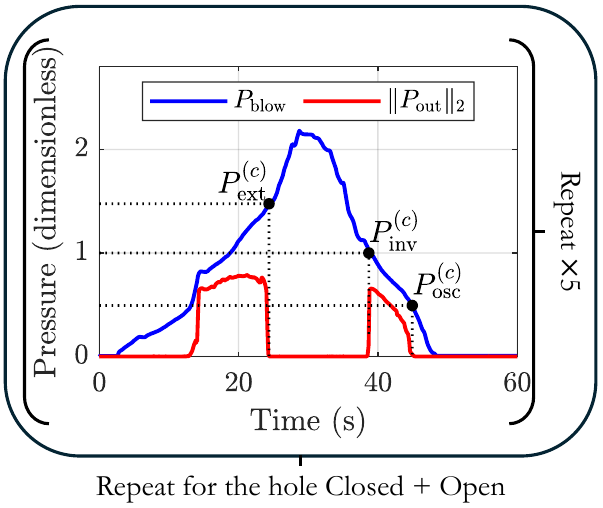}
	\caption{Experimental protocol for the blowing pressure ramps. 
	The graph shows for the closed hole, the evolution of the blowing pressure $P_\mathrm{blow}$ (in blue) and the amplitude of the external acoustic pressure recorded by the microphone $\|P_\mathrm{out}\|_2$  (in red), with respect to time.
	The three threshold values $P_\mathrm{osc}^{(c)}$, $P_\mathrm{ext}^{(c)}$, $P_\mathrm{inv}^{(c)}$ are represented.
	}
	\label{fig:schema_ramp}
\end{figure}

For a fixed embouchure, increasing and decreasing blowing pressure ramps are performed, both with the register hole closed and open.
The protocol is summarized in Figure \ref{fig:schema_ramp}.
During the \textit{crescendo} phase, the blowing pressure $P_\mathrm{blow}$ increases monotonically over a duration of 30~s.
During the \textit{diminuendo} phase, $P_\mathrm{blow}$ decreases monotonically over 30~s.
Three blowing pressure thresholds are measured.
\begin{itemize}
	\item $P_\mathrm{osc}^{(c)}$ or $P_\mathrm{osc}^{(o)}$: minimum blowing pressure that allows self-sustained oscillations.
	\item $P_\mathrm{ext}^{(c)}$ or $P_\mathrm{ext}^{(o)}$: maximum blowing pressure that allows self-sustained oscillations.
	\item $P_\mathrm{inv}^{(c)}$ or $P_\mathrm{inv}^{(o)}$: blowing pressure at which oscillations restart in the \textit{diminuendo} phase.
\end{itemize}
For each condition (hole closed or open), five blowing pressure ramps are carried out.
The three threshold values $P_\mathrm{osc}, P_\mathrm{ext}, P_\mathrm{inv}$ are averaged over the five measurements.

\subsection{Hole openings}\label{sec:exp:holeopenings}

Hole openings are performed to assess the blowing pressure range for which the second register (R2$^{(o)}$) can be reliably played when the hole is opened.
This blowing pressure range is characterized by four threshold values $$P^\mathrm{I}<P^\mathrm{II}<P^\mathrm{III}<P^\mathrm{IV},$$  defined hereafter.
\begin{itemize}
\item  $P^\mathrm{I}$: maximum  blowing pressure lower than $P^\mathrm{II}$ that never leads to R2$^{(o)}$ when the register hole is opened. 
\item  $P^\mathrm{II}$: minimum  blowing pressure that always leads to R2$^{(o)}$ when the register hole is opened. 
\item  $P^\mathrm{III}$: maximum  blowing pressure that always leads to R2$^{(o)}$ when the register hole is opened. 
\item  $P^\mathrm{IV}$: minimum  blowing pressure greater than $P^\mathrm{III}$ that never leads to R2$^{(o)}$ when the register hole is opened. 
\end{itemize}

\begin{figure}[H]
	\centering
	\includegraphics[width=.65\textwidth]{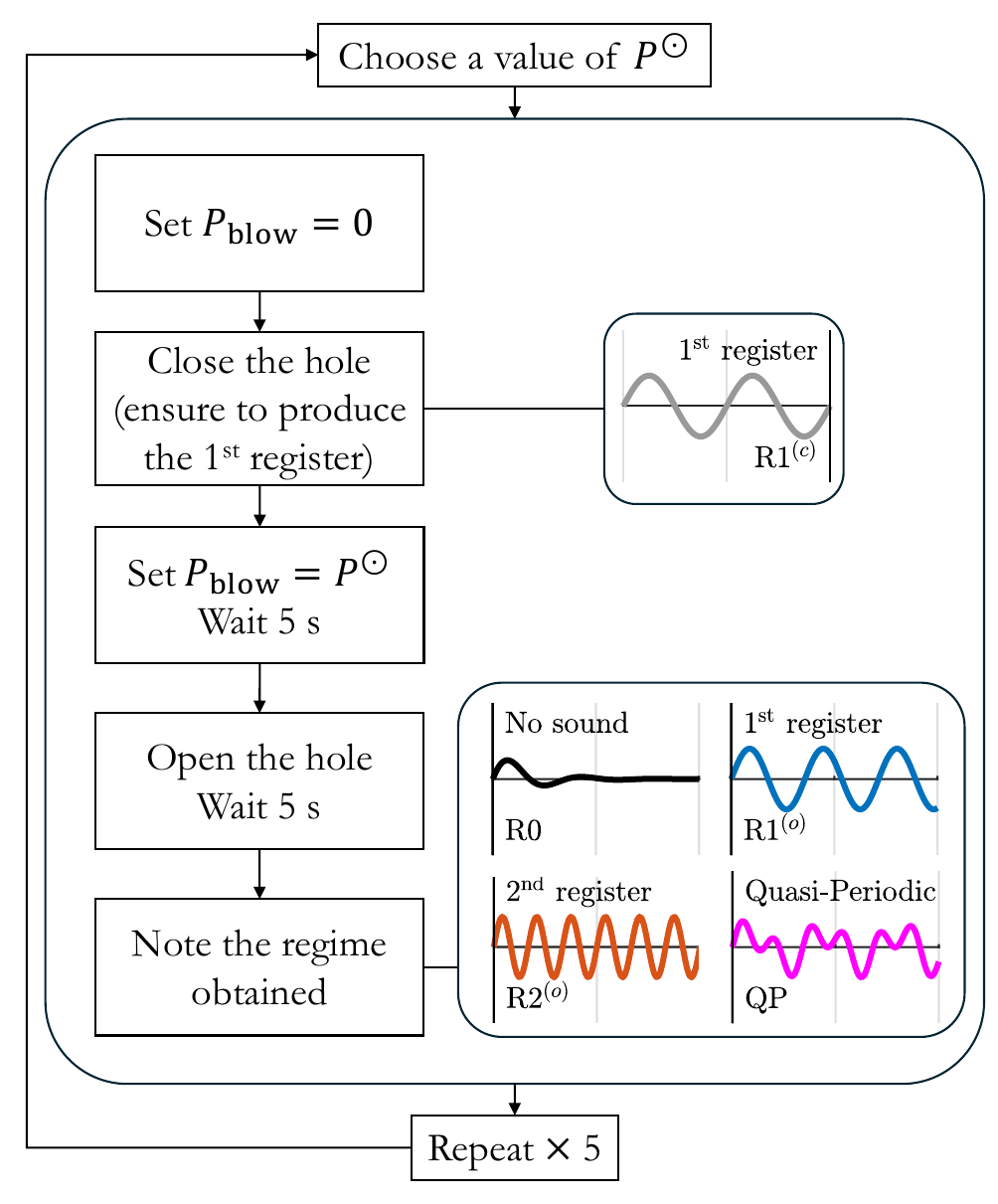}
	\caption{Experimental protocol for the hole openings procedure. The four threshold values $P^\mathrm{I}, P^\mathrm{II}, P^\mathrm{III}$ and $P^\mathrm{IV}$, are determined iteratively by opening the hole five times for a selected blowing pressure, $P^\odot$.
	$P^\odot$ is then increased until $P^\mathrm{IV}$ is found.}
	\label{fig:schema_open}
\end{figure}

To measure the value of a given threshold, the following method is employed, also described in Figure~\ref{fig:schema_open}.
For a tested value of the blowing pressure, noted $P^\odot$, five hole openings are realized.
After each opening, the type of register obtained is noted.
The following notations are used, as shown on Figure~\ref{fig:schema_open}: R0 for the equilibrium (no sound), R1$^{(o)}$ for a first register, R2$^{(o)}$ for the second register, QP for a quasi-periodic regime.
If the second register R2$^{(o)}$ is obtained five times out of five, the blowing pressure tested is considered as reliable to play twelfths.
If, however, R2$^{(o)}$ is never obtained, the second register is considered unplayable for the pressure tested when the hole is opened.

Results are presented in Section~\ref{sec:results:exp}.

\section{Numerical model}\label{sec:model}
\subsection{Digital resonators}\label{sec:model resonator}

The digital resonator is presented on Figure \ref{fig:2} and has the same dimensions as the simplified clarinet used for the experiment.
It is composed of a first tube of length $L_1=132$~mm, radius $R=7.5$~mm and cross-section $S=\pi R^2$.
The characteristic impedance of plane waves  propagating through the tube is $Z_c=\rho_0 c_0/S$ where $\rho_0=1.23~\mathrm{kg\cdot m^{-3}}$ and $c_0=343~\mathrm{m\cdot s^{-1}}$.
The acoustic field in the first tube is described by the pressure at the left extremity $p_{in}$, and at the right extremity $p_1$.

The tube is branched to a side hole of length $L_h$~$=$~$10$~mm, radius $R_h=1.5$~mm, cross-section $S_h$~$=$~$\pi R_h^2$, and characteristic impedance $Z_{ch}= \rho_0 c_0 / S_h$.
The acoustic field in the side hole is described by the pressure at the bottom of the hole $p_{hb}$ and at the top of the hole $p_{ht}$.

A second tube of length $L_2=166$~mm and cross-section $S$ is branched downstream the side hole.
The acoustic field in this tube is described by the pressure at the left extremity $p_2$ and by the pressure at the right extremity $p_{end}$.

\subsection{Viscothermal losses}
Viscothermal losses are introduced through the complex wavenumber $\Gamma_i(s)$, where $s$ is the Laplace variable and $i=\{1,2,h \}$ refers to the index of the tube considered. 
The function $G_i(s)$ is defined, such that
\begin{align*}
G_i(s) &= e^{-\Gamma_i(s) L_i} = \lambda_i e^{-\epsilon_i \sqrt{s}} e^{-\tau_i s},
\end{align*}
with
\begin{align*}
\lambda_i &= e^{-({\alpha_2 \ell_v L_i})/{R_i^2}}, &
\epsilon_i &= \frac{\alpha_1 L_i}{R_i}\sqrt{\frac{2 \ell_v}{c_0}}, & \tau_i &= \frac{L_i}{c_0}, 
\end{align*}
where $\alpha_1=1.044$,  $\alpha_2=1.080$,  and $\ell_v= 4 \cdot 10^{-8}~\mathrm{m}$ [Chap.\ 5.5 of Chaigne and Kergomard (2016)]\cite{bible2016}.

In practice, $G_i(s)$ are approximated by a first-order low-pass filter and a delay $\tilde{G}_i(s)$, following the work from Guillemain \textit{el al.}\ (2005)\cite{guillemain2005real}.
Fractional delays $\tau_i$ are also taken into account through the order 1 filters proposed by Laakso \textit{et al}.\ (1996)\cite{laakso1996splitting}.

\subsection{Forward and backward-propagating pressure waves}\label{sec: model eqs}
In the following, time-domain variables are written in small letters ({e.g.}\ $p_2^+(t)$), and frequency-domain variables are written in capital letters ({e.g.}\ $P_2^+(s)$).

The propagation of the acoustic waves in the resonator is described through the forward and backward-propagating acoustic pressures $p^+$ and $p^-$.
They are related to the acoustic pressure and flow $(p, u)$ through the relationships:
\begin{align*}
p&= p^+ + p^-, & u &= \frac{p^+ - p^-}{Z},
\end{align*}
where $Z=Z_c$ in the main tube of cross-section $S$, and $Z= Z_{ch}$ in the side hole.

Since the tubes $L_1$, $L_2$ and $L_h$ are all cylindrical, the acoustic field can be described as transmission lines equations in the frequency domain, following Figure~\ref{fig:2}.
For the tube of length $L_1$:
\begin{align}
P_1^+ &= G_1 P_{in}^+, & P_{in}^- &= G_1 P_{1}^- .
\end{align}
For the tube of length $L_2$:
\begin{align}
P_{end}^+ &= G_2 P_{2}^+, & P_{2}^- &= G_2 P_{end}^- .
\end{align}
For the tube of length $L_h$:
\begin{align}
P_{ht}^+ &= G_h P_{hb}^+, & P_{hb}^- &= G_h P_{ht}^- .
\end{align}

\subsection{Boundary conditions}\label{sec:bcs}
The boundary conditions in the tube are described hereafter.

\subsubsection{Radiation}
First, radiation from the open end is neglected: the pressure $p_{end}$ is written consequently as
\begin{equation}
 p_{end} = 0.
\end{equation}

\subsubsection{Hole junction}
Secondly, since the register hole has a small diameter and a long chimney length, the series impedances of the hole can be neglected [section 3.3.2.2 of Debut \textit{et al}.\ (2005)]\cite{debut_analysis_2005}.
The boundary conditions applied to the acoustic pressure and flow at the bottom of the hole are therefore given by:
\begin{align}
p_1 &= p_2, \label{eq:p1p2}  \\ 
p_2 &= p_{hb}, \label{eq:p2ph}  \\ 
u_1 &= u_2 + u_{hb}. \label{eq:flowHole}
\end{align}

\subsubsection{Flow crossing the reed channel}
The next boundary condition involves $p_{in}$ and comes from the nonlinear characteristics of the flow entering the resonator. 
In this relationship, the acoustic flow $u_{in}$ depends on the difference between the blowing pressure of the musician $p_m$ and the pressure at the input of the instrument $p_{in}$.
By assuming that the jet experiences total turbulent dissipation \cite{wilson_operating_1974} and modeling  the reed as a massless, undamped spring \cite{ollivier2005idealized}, the nonlinear characteristics is defined as \cite{dalmont2003nonlinear}:
\begin{equation}\label{eq:uin}
\hat{u}_{in} = \zeta [\hat{p}_{in} - \gamma +1]^+ \text{sgn}(\gamma - \hat{p}_{in}) \sqrt{|\gamma - \hat{p}_{in}|}, 
\end{equation}
where the function $[x]^+$ returns the positive-part of $x$, i.e.\ $ [x]^+ = (x + |x|)/2$.
The dimensionless blowing pressure is given by $\gamma= p_m/P_M$, where $P_M$ is the minimum pressure needed to close the reed channel in a quasi-static regime. 
Typical values of $P_M$ are in the range $P_M\in[4, 8]$~kPa, according to Dalmont and Frapp\'e (2007)\cite{dalmont_oscillation_2007}, and Atig \textit{el al}.\ (2004)\cite{atig2004saturation}.
The parameter $\zeta$ represents the embouchure, with common values for the clarinet between 0.05 and 0.4 \cite{dalmont_oscillation_2007}.
The dimensionless quantities are defined as
\begin{align*}
\hat{p}_{in}&=p_{in} / P_M, & \hat{u}_{in} &=  u_{in}Z_c/ P_M.
\end{align*}

In Eq.\ \eqref{eq:uin}, the dynamics of the reed are neglected to obtain a direct relationship between $p^+_{in}$ and $p^-_{in}$. 
This relationship is given in Taillard \textit{et al}.\ (2010)\cite{taillard2010iterated} and is detailed in the Appendix of Bergeot \textit{et al.}\ (2014)\cite{bergeot2014response}.
It is expressed as:
\begin{align}\label{eq:Raman}
\hat{p}_{in}^+ = f_{\gamma \zeta}(\hat{p}_{in}^-)= \gamma - X[\gamma - 2 \hat{p}_{in}^-] - \hat{p}_{in}^-,
\end{align}
where the function $X$ is defined in Appendix A of Taillard \textit{et al}.\ (2010) \cite{taillard2010iterated}.

\subsubsection{Localized nonlinear losses in the register hole}
Localized nonlinear losses in the register hole are modeled using the following boundary condition for $p_{ht}$:
\begin{equation}\label{eq:bcnl pv}
p_{ht}(t) = \rho_0 C_\text{nl} v_{ht}(t)|v_{ht}(t)|,
\end{equation}
where $C_\text{nl}>0$ is the nonlinear loss coefficient, which depends on the roundness of the edges of the hole \cite{atig2004saturation}, and $v_{ht}$ is the acoustic speed at the top of the side hole. 
An explicit relationship between $p^+_{ht}$ and  $p^-_{ht}$ is given in Szwarcberg \textit{et al}.\ (2025)\cite{szwarcberg2025oscillation}:
\begin{equation}
p_{ht}^-(t) = r_\mathrm{nl}\left[p_{ht}^+(t)\right],
\end{equation}
where
\begin{align}\label{eq:bcnl}
r_\mathrm{nl}(x) = x \left(1- \frac{4}{1+ \sqrt{1+K_\mathrm{nl}|x|}} \right),
\end{align}
with $K_\mathrm{nl}=8C_\mathrm{nl}/(\rho_0 c_0^2)$.
For $K_\mathrm{nl}=0$, we get $r_\mathrm{nl}(x)$~$=$~$-x$, which corresponds to an open hole boundary condition.
As $K_\mathrm{nl} \to \infty$, $r_\mathrm{nl}(x)=x$, meaning the hole is closed.

In a dimensionless form, $r_\mathrm{nl}$ is rewritten as $\hat{p}_{ht}^-$~$=$~$\hat{r}_\mathrm{nl}\left[\hat{p}_{ht}^+\right]$, where
\begin{align}\label{eq:bcnl adim}
\hat{r}_\mathrm{nl}(x) = x \left(1- \frac{4}{1+ \sqrt{1+\hat{K}_\mathrm{nl}|x|}} \right),
\end{align}
with $\hat{K}_\mathrm{nl}= P_M K_\mathrm{nl}=0.1$, assuming a low $P_M$ and a hole with sharp edges.

\subsection{Implementation of the model}
All details on the model implementation are provided in the Supplementary Material (Sec.~1).

\subsection{Extraction of the modal acoustic pressure}
Modal acoustic pressures are useful to visualize the limit cycles of the different oscillating regimes.
However, they are not directly accessible through waveguide modeling. 
Filtering is applied \textit{a posteriori}, using  the modal decomposition of the input impedance $Z_{in}=P_{in} /U_{in}$:
\begin{equation}
Z_{in}   = Z_c\sum_n \frac{C_n}{s-s_n} + \frac{\mathrm{conj}(C_n)}{s-\mathrm{conj}(s_n)},
\end{equation}
where $C_n$ and $s_n$ are the complex residues and poles, computed through the residues theorem from the analytic definition of the input impedance\cite{monteghetti2018analysis}.
In particular, the modal frequencies are given by $f_n=\Im(s_n)/(2\pi)$.
The $n$-th modal acoustic pressure at the input $p_n$ is defined through the following ODE:
\begin{equation}
\dot p_n(t) = Z_c C_n u_{in}(t) + s_n p_n(t),
\end{equation}\label{eq:modal}
where $\dot p_n=\partial_t p_n$ and $u_{in}=(p_{in}^+  - p_{in}^-)/Z_c$.
Modal acoustic pressures can then be computed by filtering $u_{in}$ with an IIR filter.
\newpage
\section{Simulations}
The control parameter space ($\gamma, \zeta$) is first mapped out to find the ranges for which the first register is stable when the hole is closed (R1$^{(c)}$), as well as the range for which the second register is stable when the hole is open (R2$^{(o)}$).  

Multiple hole openings are then performed for constant control parameters to assess the playability of the  register jumps.

\subsection{Cartography of the playing range of the different registers }\label{sec:sim:ramps}
\begin{figure}[H]
	\centering
	\includegraphics[width=\textwidth]{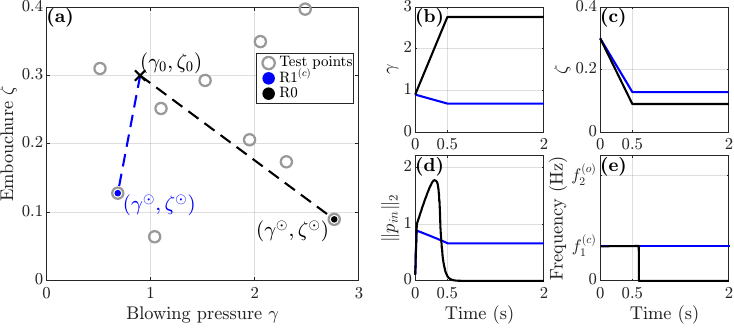}
	\caption{Example of a cartography of the control parameter space $(\gamma, \zeta)$ by Latin Hypercube sampling (a).
	The time evolution of $\gamma$ (b), $\zeta$ (c), the amplitude of the acoustic pressure $\|p_{in}\|_2$ (d) and the playing frequency (e) is also shown for two different target control parameters.
	In this example, the hole is closed.
	}
	\label{fig:schema_carto}
\end{figure}

This section adopts a regime cartography method inspired by Colinot \textit{et al.\ }(2025)\cite{colinot2025cartography}.
It is summarized in Figure~\ref{fig:schema_carto}.
In the control parameter space ($\gamma, \zeta$), the regions in which two regimes are stable are determined: the first register for the closed hole R1$^{(c)}$, and the second register for the open hole R2$^{(o)}$.
The closed and open hole cases correspond to values of the nonlinear losses coefficient of $\hat K_\mathrm{nl}=+\infty$ and $\hat K_\mathrm{nl}=0.1$, respectively.

Each simulation is run with a set of control parameters $\gamma^\odot$ and $\zeta^\odot$. 
The control parameter space $(\gamma, \zeta)$ is mapped by $N_{\gamma \zeta}=10^4$ Latin hypercube samples (each sample is alone in each axis-aligned
hyperplane containing it \cite{mckay2000comparison}).
They are distributed in the range $\gamma \in [0.3, 3]$ and $\zeta \in [0.05, 0.4]$.
The boundaries in $\zeta$ correspond to clarinet playing conditions \cite{dalmont_oscillation_2007, fritz2004clarinette}.

For R1$^{(c)}$ and R2$^{(o)}$, to enable the model to play a stable register for a target blowing pressure and embouchure $(\gamma^\odot, \zeta^\odot )$, the control parameters are first linearly interpolated from $(\gamma_0, \zeta_0)=(0.9,0.3)$ to $(\gamma, \zeta)=(\gamma^\odot, \zeta^\odot)$ for a duration $t_\mathrm{var}=0.5$~s.
The values of $\gamma_0$ and $\zeta_0$ were identified as suitable starting points based on preliminary tests.
From time $t>t_\mathrm{var}$, the control parameters are kept at their target values until the end of the simulation at time $t_\mathrm{max}=2$~s.
The register played is then determined.

Multistability zones are assessed using this method. 
From the nonlinear characteristics of the flow through the reed channel [Eq.~\eqref{eq:uin}], it follows that the equilibrium state (no sound, denoted R0) is stable when $\gamma > 1$ \cite{dalmont2005analytical}.

\subsection{Hole openings}\label{sec:sim:holeopenings}

For constant control parameters and from a first register on the closed hole (R1$^{(c)}$), the hole is instantaneously opened.
$N_o=200$ hole openings are performed at different times $t_\mathrm{open} \in [T_\mathrm{open}- \frac{1}{2} T_\mathrm{play}^{(c)}, T_\mathrm{open}+ \frac{1}{2} T_\mathrm{play}^{(c)}]$, with $T_\mathrm{open}=1.0$~s and $T_\mathrm{play}^{(c)}=1/f_\mathrm{play}^{(c)}$.
$f_\mathrm{play}^{(c)}=274$~Hz is the frequency of the input pressure signal for a first register with the hole closed.
Hence, a limit cycle of R1$^{(c)}$ can be fully sampled.
After the opening, simulations continue until time $t_\mathrm{max}=4.0~$s.
The proportion of each register obtained for the $N_o$ openings is computed.

By dividing the control parameter space into $\zeta$ slices, four threshold values are measured for each slice: $\gamma^\mathrm{I}, \gamma^\mathrm{II}, \gamma^\mathrm{III}, \gamma^\mathrm{IV}$.
These values are equivalent to $P^\mathrm{I}, P^\mathrm{II}, P^\mathrm{III}, P^\mathrm{IV}$ under a dimensionless formalism.
In practice, the thresholds are sought from $\zeta=0.05$ to $\zeta=0.40$, with a step size $\Delta \zeta=0.05$.

Finally, to extend the analysis of the basins of attraction of the regimes associated with the open hole, initial conditions slightly away from the initial limit cycle of R1$^{(c)}$ are generated by adding uniform white noise to $\gamma$ and $\zeta$ for $t < t_\mathrm{open}$.
Different perturbations are tested, with amplitudes between $0~\%$ and $30~\%$ of the value of the control parameter.

The control parameters are kept steady (without noise) for $t\geq t_\mathrm{open}$ to ensure that the basins of attraction of the regimes relative to the open hole are unchanged.

\section{Results}

\subsection{Experiment}\label{sec:results:exp}

\begin{figure}[H]
	\centering
	\includegraphics[height=.42\textwidth]{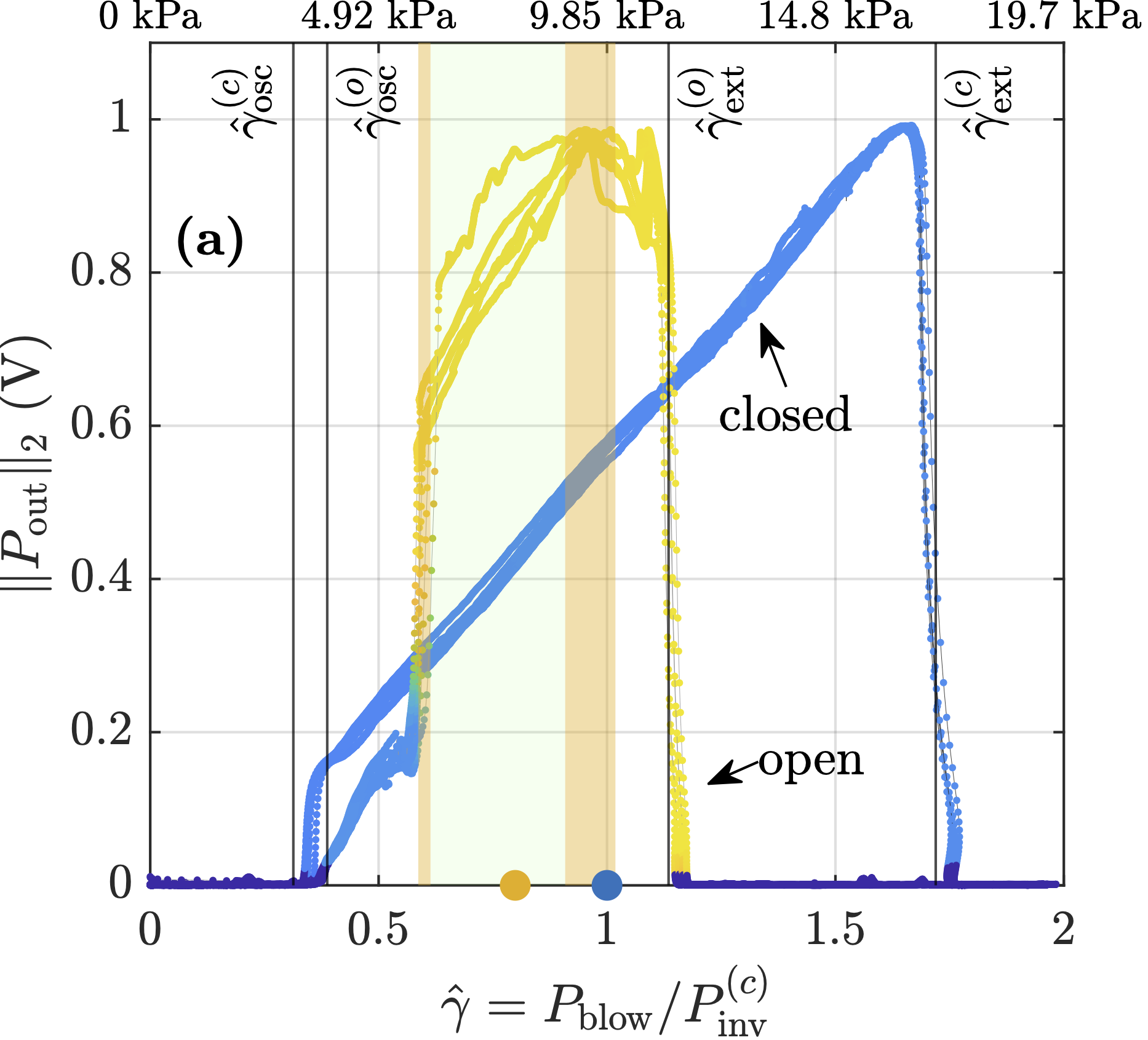}
	\includegraphics[height=.42\textwidth]{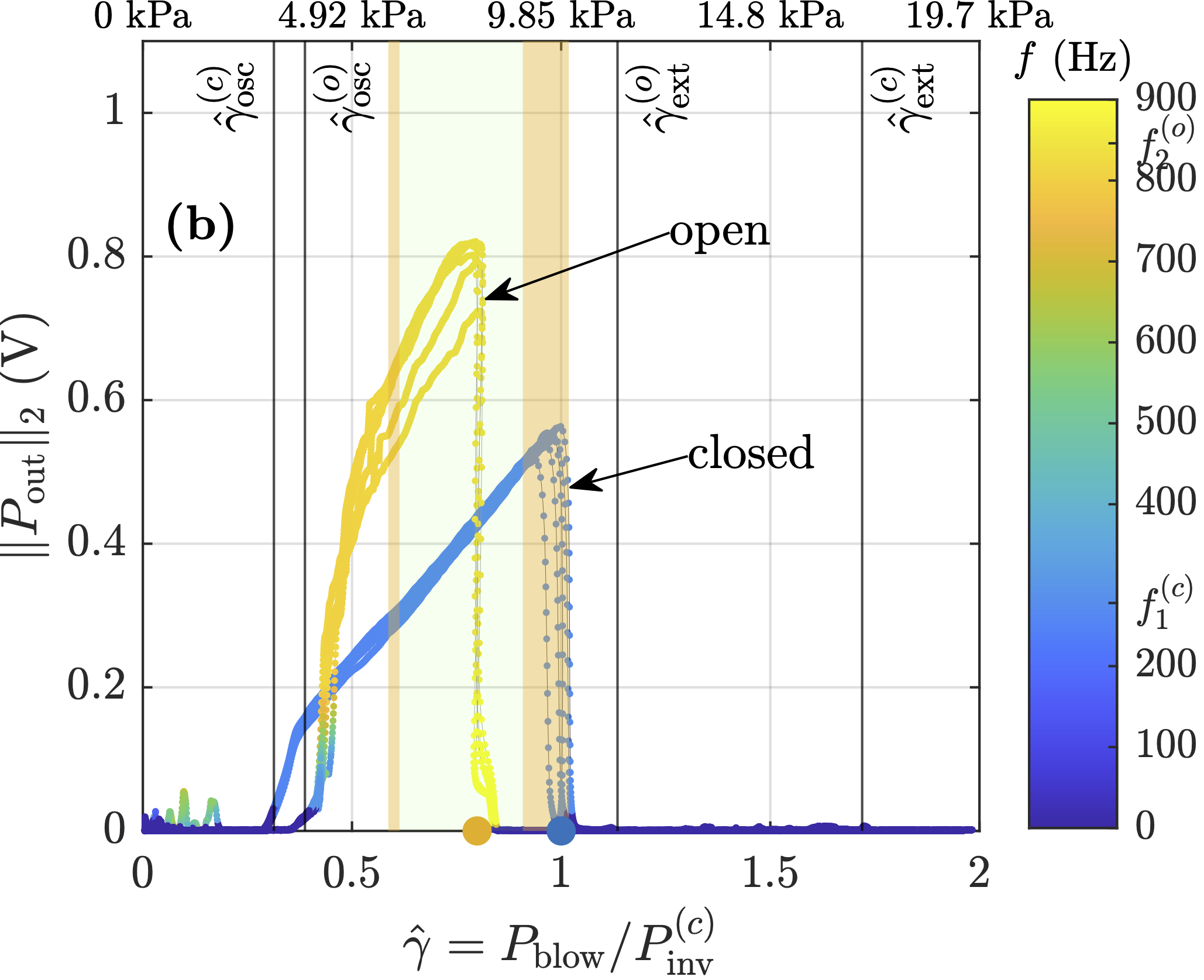}
	\caption{Evolution of the amplitude of the external acoustic pressure $P_\mathrm{out}$ when the scaled blowing pressure $\hat \gamma$ increases (a) or decreases (b) monotonically, for a constant embouchure, and for the hole closed and open. 
	The measured values of $P_\mathrm{blow}$ are displayed at the top of the graphs.
	Five measurements are represented for each case.
	The color of the curves is indexed on the playing frequency. 	
	The blue and yellow dots on the x-axis show the values of $\hat \gamma_\mathrm{inv}^{(c)}$ and $\hat \gamma_\mathrm{inv}^{(o)}$ respectively.	
	Colored surfaces in the background show the ranges of $\hat \gamma_\mathrm{blow}$ where the second register is reached with a given probability when the hole is opened.
	Green: $100~\%$.
	orange: between $0~\%$ and $100~\%$.}
	\label{fig:exp}
\end{figure}

Figure~\ref{fig:exp}(a) shows how the amplitude of the external acoustic pressure $P_\mathrm{out}$ evolves during a \textit{crescendo} ($P_\mathrm{blow}$ increases from 0 to 20~kPa over 30~s). 
Figure~\ref{fig:exp}(b) shows the \textit{diminuendo} phase ($P_\mathrm{blow}$ decreases from 20~kPa to 0~kPa over 30~s). 
Five measurements are made for each configuration (closed, open).
To facilitate comparison with the numerical results, the x-axis is rescaled using the variable transformation $$\hat \gamma = P_\mathrm{blow}/ P_\mathrm{inv}^{(c)}.$$  
The measured values of $P_\mathrm{blow}$ are indicated at the top of the graphs in Figure~\ref{fig:exp}, and in Supplementary~Figure~2.

For the closed hole, the playing frequency is close to the first modal frequency $f_1^{(c)}$, which corresponds to the first register R1$^{(c)}$. 
For the open hole, two different oscillating regimes are played.
First, Figure \ref{fig:exp} shows a stable first register R1$^{(o)}$ for $P_\mathrm{blow}\in[3.7, 5.8]$~kPa (in blue).
Secondly, in the \textit{crescendo} phase (Fig.~\ref{fig:exp}(a)), a stable second register R2$^{(o)}$ is produced for $P_\mathrm{blow} \in [5.8, 11.1]$~kPa (in yellow).
However, in the \textit{diminuendo} phase (Fig.~\ref{fig:exp}(b)), R2$^{(o)}$ is also stable for $P_\mathrm{blow}\in [4.3, 7.9]$~kPa.
Thus, R2$^{(o)}$ is stable for $P_\mathrm{blow}\in [4.3, 11.1]$~kPa.

For both the open and the closed hole, the three thresholds $P_\mathrm{osc}, P_\mathrm{ext}$ and $P_\mathrm{inv}$ are measured.
Their values are given in Table~\ref{tab:1}.
Note that the value of $P_\mathrm{blow}$ at which the oscillations start during the \textit{crescendo} phase is greater than the value of $P_\mathrm{blow}$ at which the oscillations stop during the \textit{diminuendo} phase.
This discrepancy is due to bifurcation delay \cite{bergeot2013prediction}.

Furthermore, it is surprising to observe that $P_\mathrm{inv}^{(o)}$ differs from $P_\mathrm{inv}^{(c)}$.
In theory, $P_\mathrm{inv}=P_M=K_r H_0,$ where $K_r$ is the stiffness of the reed and $H_0$ is the height of the reed at rest.
$H_0$ remains constant during the experiment since the chamber of the artificial player system is never opened.
$K_r$ may decrease with time due to fatigue of the reed and therefore reduce the value of $P_\mathrm{inv}$ after successive repetitions.
This hypothesis is rejected by repeating the experiment, alternating ramps with the hole open and closed, and observing the same results.
Simulations with a more complex physical model \cite{taillard2018modal} highlight the significant role of the reed flow \cite{chabassier2022control} and the reed damping \cite{wilson_operating_1974} on the discrepancy between the closed and the open hole cases.
They are shown in the Supplementary Figures 3 and 4.

\begin{table}[H]
		\centering
		\caption{
Mean values of the experimental thresholds and their associated standard deviation, over five measurements.
		}
		\label{tab:1}
		\begin{tabular}{lcccccc}
		\hline
		Threshold value & $P_\mathrm{osc}^{(c)}$ & $P_\mathrm{inv}^{(c)}$ & $P_\mathrm{ext}^{(c)}$ & $P_\mathrm{osc}^{(o)}$ & $P_\mathrm{inv}^{(o)}$& $P_\mathrm{ext}^{(o)}$ \\
		Mean (kPa) & 3.08 & 9.85 & 16.9 & 3.82 & 7.87 & 11.2\\
		Standard deviation (kPa) & 0.027 & 0.190 & 0.202 & 0.031 & 0.033 & 0.069\\
		\hline
		\end{tabular}
\end{table}

By combining information on the stability of the different regimes produced during the \textit{crescendo} and the \textit{diminuendo} phases, the following multistability regions can be determined.
\begin{itemize}
\item $P_\mathrm{blow}\in [4.3, 5.8]$~kPa: multistability between R2$^{(o)}$ and R1$^{(o)}$.
\item $P_\mathrm{blow}\in [P_\mathrm{inv}^{(o)}, P_\mathrm{ext}^{(o)}]$ or $P_\mathrm{blow}\in [7.9, 11.1]$~kPa: multistability between R2$^{(o)}$ and R0.
\end{itemize}

In the two regions where R2$^{(o)}$ is multistable with another regime, if clarinetists play in the first register with the hole closed and then open the register hole, they may end up in either R2$^{(o)}$ or another regime.
The hole opening procedure described in Section~\ref{sec:exp:holeopenings} is used to assess how  the probability of playing R2$^{(o)}$ evolves with $P_\mathrm{blow}$.
The four threshold values characterizing this probability are listed in Table~\ref{tab:threshOpen}.

Within the green region displayed on Figure \ref{fig:exp}, R2$^{(o)}$ is always reached. 
In the orange regions around $P_\mathrm{blow}=6.0$~kPa and $9.5$~kPa, different regimes can be observed (R0, R1$^{(o)}$, R2$^{(o)}$ or QP) when opening the hole.
Thus, we notice that the blowing pressure range in which R2$^{(o)}$ can be played after opening the register hole from R1$^{(c)}$ is narrower than the blowing pressure range where R2$^{(o)}$ is stable.

These experimental observations are reproduced digitally, following the protocol defined in Sections~\ref{sec:sim:ramps} and \ref{sec:sim:holeopenings}.
A particular focus is given to the transition (orange) regions.

\begin{table}[h!]
\centering
\caption{Experimental values of the four thresholds characterizing the probability of playing the second register when opening the hole.}
\label{tab:threshOpen}
\begin{tabular}{lcccc}
\hline
Threshold & $P^\mathrm{I}$ & $P^\mathrm{II}$ & $P^\mathrm{III}$& $P^\mathrm{IV}$ \\
Value (kPa) & $5.8$ & $6.0$ & $8.9$ & $10.0$ \\
\hline
\end{tabular}
\end{table}

\newpage
\subsection{Simulations}
\label{sec:results:sim}

\subsubsection{Reproduction of the experiment}

\begin{figure}[H]
		\centering
		\includegraphics[height=.4\textwidth]{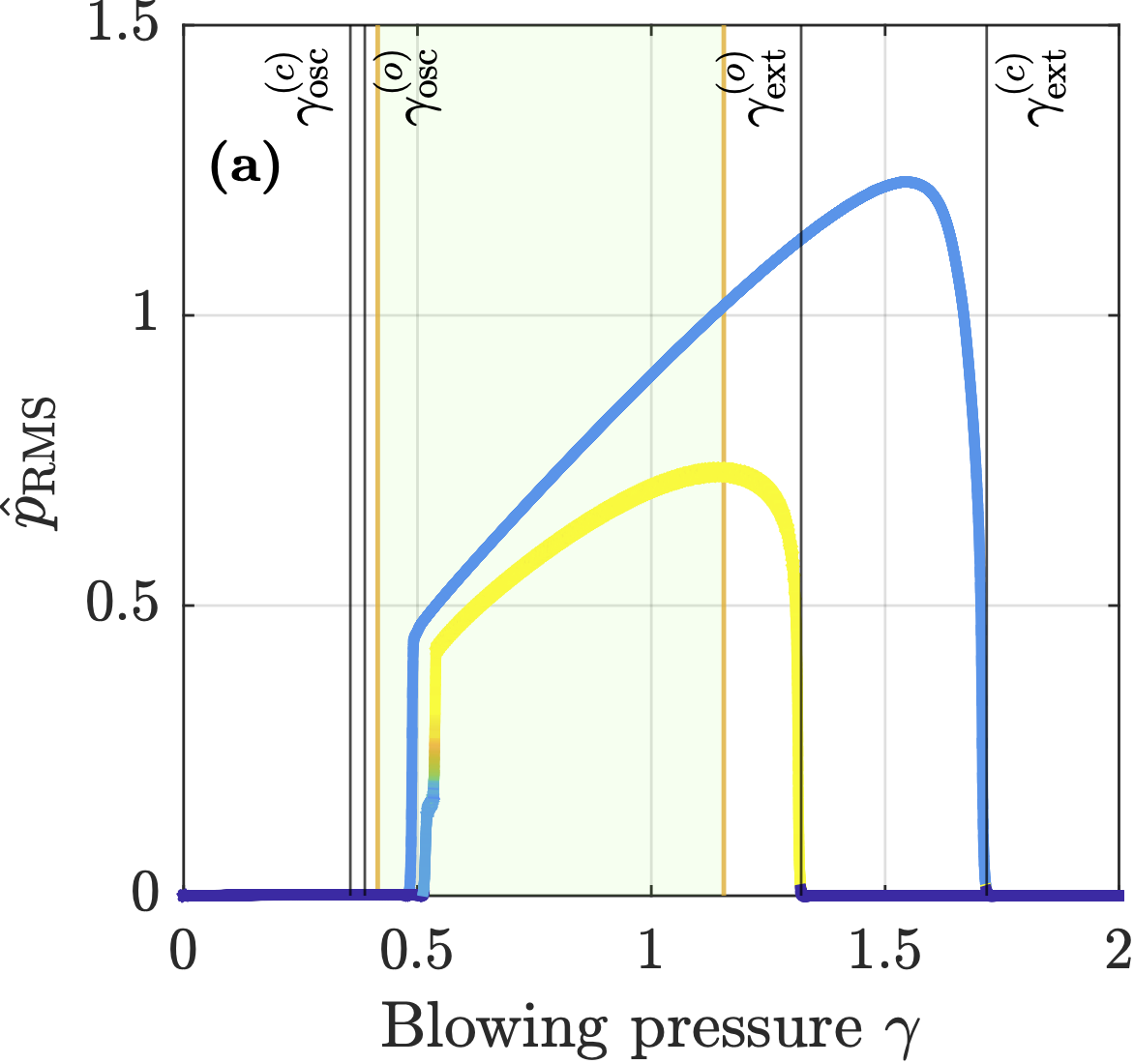}
		\includegraphics[height=.4\textwidth]{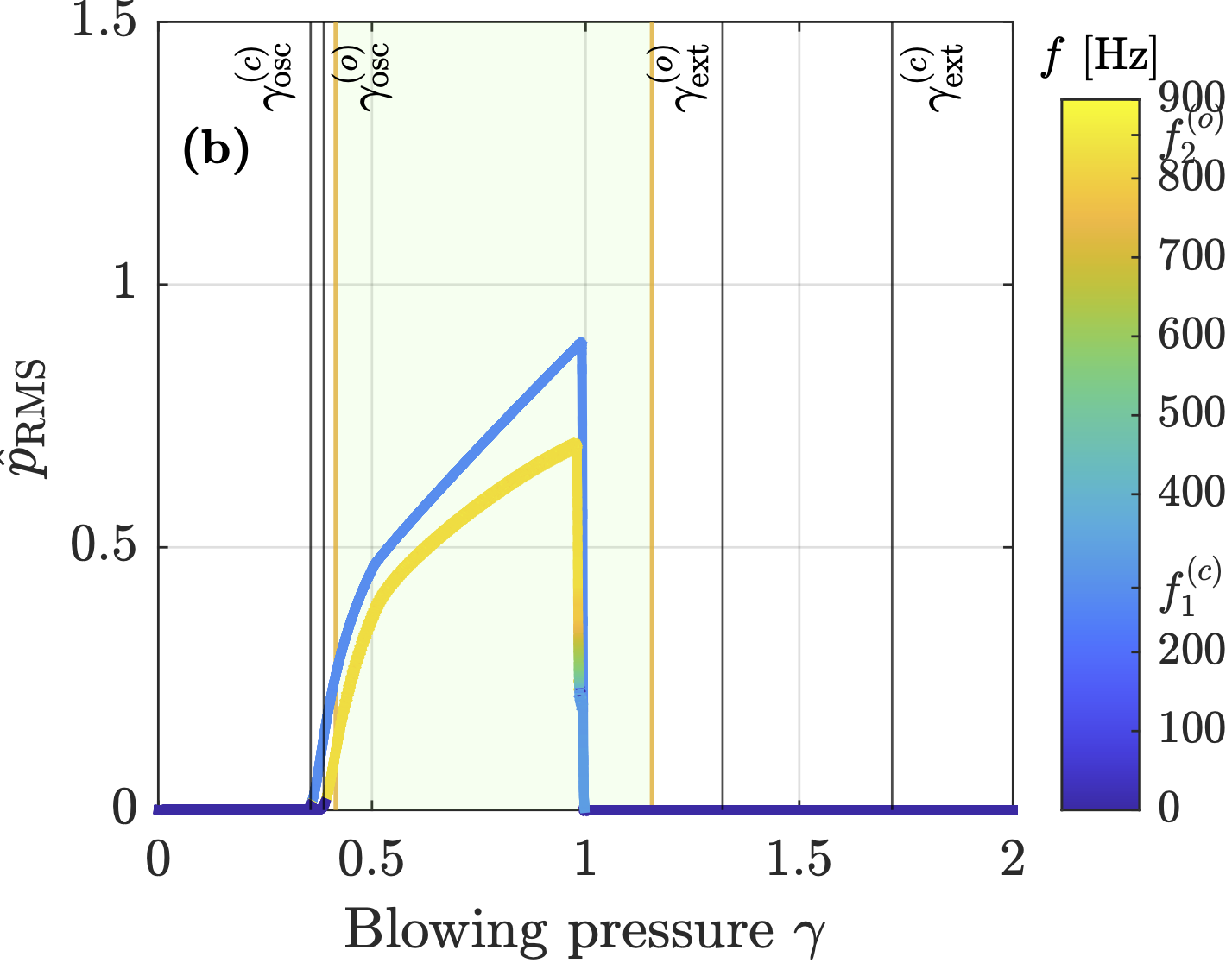}
		\caption{
		Evolution of the RMS amplitude of the input acoustic pressure $\hat p_\mathrm{in}$, as $\gamma$ increases (a) linearly from $\gamma=0$ to $\gamma=2$ during 30~s, then decreases (b) during the same duration.
		Simulations with the hole closed and open are represented.
		The value of the embouchure parameter is $\zeta=0.17$. 
		Colored surfaces in the background have the same meaning as Figure~\ref{fig:exp}. }			
		\label{fig:simCD}
\end{figure}

As a preliminary exploration, the experiment is reproduced digitally.
Simulations are performed, where the blowing pressure $\gamma$ increases linearly from $0$ to $2$ during 30~s (Figure~\ref{fig:simCD}(a)), then decreases from $2$ to $0$ during the same duration (Figure~\ref{fig:simCD}(b)). 
The simulations are carried out for both the closed (blue) and open (yellow) register hole configurations. 
A value of $\zeta = 0.17$ allows the model to reproduce the extinction threshold for the closed hole, $\gamma_\mathrm{ext}^{(c)} = 1.72$. 
The comparison of the dimensionless thresholds is provided in Table~\ref{tab:comparison_thresh}. 
These simulations reproduce the main qualitative features of the experimental results. 
In particular, during the crescendo phase with the open hole, the model captures the transition through an intermediate first-register regime R1$^{(o)}$ around $\gamma = 0.5$, before overblowing to R2$^{(o)}$.

The main difference between the model and the experiment lies on the discrepancy on the inverse threshold between the open and closed hole cases, which is clearly observed experimentally but not captured by the present model. 
Moreover, for the open hole configuration, the stability range of the oscillating regimes (R1$^{(o)}$ and R2$^{(o)}$) is larger in the simulations ($\gamma \in [0.39, 1.32]$) than in the experiments ($\hat{\gamma} \in [0.39, 1.14]$).

\begin{table}[H]
		\centering
		\caption{
		Dimensionless experimental and numerical thresholds.
		The experimental dimensionless thresholds $\hat \gamma$ are computed by dividing the mean scaled values by the mean value $P_\mathrm{inv}^{(c)}=9.85$~kPa (Table~\ref{tab:1}).
		}
		\label{tab:comparison_thresh}	
		\begin{tabular}{lcccccc}
		\hline 
		Experimental  & $\hat \gamma_\mathrm{osc}^{(c)}$ & $\hat \gamma_\mathrm{inv}^{(c)}$ & $\hat \gamma_\mathrm{ext}^{(c)}$ & $\hat \gamma_\mathrm{osc}^{(o)}$ & $\hat \gamma_\mathrm{inv}^{(o)}$& $\hat \gamma_\mathrm{ext}^{(o)}$ \\
		~ & 0.31 & 1.0 & 1.72 & 0.39 & 0.80 & 1.14\\
		Numerical & $\gamma_\mathrm{osc}^{(c)}$ & $ \gamma_\mathrm{inv}^{(c)}$ & $ \gamma_\mathrm{ext}^{(c)}$ & $ \gamma_\mathrm{osc}^{(o)}$ & $ \gamma_\mathrm{inv}^{(o)}$& $ \gamma_\mathrm{ext}^{(o)}$\\
		~ & 0.36 & 1.0 & 1.72 & 0.39 & 1.0 & 1.32 \\
		\hline
		\end{tabular}
\end{table}
	
\subsubsection{Cartography}\label{sec:results:carto}
\begin{figure}[H]
	\centering
	\includegraphics[width=.6\textwidth]{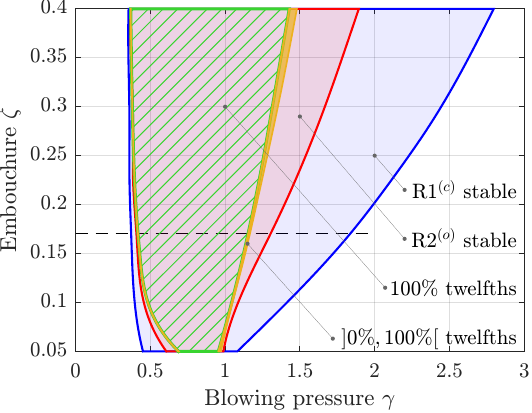}
	\caption{Simulation results: stability of the different registers in the $(\gamma,\zeta)$ plane, and playability of the twelfths.
	For control parameters located in the blue surface, the first register is stable for the closed hole case (R1$^{(c)}$).
	Within the red surface, the second register is stable for the open hole case (R2$^{(o)}$).
	Within the hatched green region, R2$^{(o)}$ is always reached when opening the hole from R1$^{(c)}$.
	Within the orange regions, the register obtained when the hole is opened depends on the timing of the opening of the hole.
	Outside of the green and orange regions, R2$^{(o)}$ is never obtained when opening the hole from R1$^{(c)}$.
	The black dashed line shows the control parameters used in Figure~\ref{fig:simCD}.
	}
	\label{fig:gz}
\end{figure}

Figure \ref{fig:gz} shows the values of the blowing pressure $\gamma$ and the embouchure parameter $\zeta$ where the first register is stable for the closed hole (R1$^{(c)}$, in blue), and the second register is stable for the open hole (R2$^{(o)}$, in red).
The equilibrium (R0, no sound) is stable for $\gamma\geq 1$.
In the green hatched region, R2$^{(o)}$ is always obtained when the register hole is opened multiple times from R1$^{(c)}$, with the protocol detailed in Section~\ref{sec:sim:holeopenings}.
This green region is surrounded by two thin orange regions.
In these regions, the regime obtained after opening the hole changes with respect to the time at which the hole is opened.
Outside the green and orange regions, R2$^{(o)}$ is never played when the register hole is opened from R1$^{(c)}$.
For high blowing pressures (blue and red surfaces for $\gamma>1$), R0 is always obtained.
For lower blowing pressures, when R2$^{(o)}$ is stable (thin red region around $\zeta<0.15$), the first register R1$^{(o)}$ is always obtained when the hole is opened.

As in the experiment, simulations show that the region where stable register jumps can be produced from R1$^{(c)}$ to R2$^{(o)}$ is narrower than the region where R2$^{(o)}$ is stable.
In addition, we notice again that for low blowing pressures, the stability region of R2$^{(o)}$ almost coincides with the region in which stable register jumps can be played.
Furthermore, Figure \ref{fig:gz} shows that when playing \textit{pianissimo} by blowing very softly, a relaxed embouchure (i.e.\ a high value of $\zeta$) may be essential to play register jumps.

A closer look at the transition regions (in orange) is carried out in the next section. 

\subsubsection{Multistable transitions at high blowing pressure}\label{sec:results:multi}

We now focus on the high blowing pressure region, specifically for values of $\gamma$ in the interval $[\gamma^\mathrm{III}, \gamma^\mathrm{IV}]$. In this range, the regime that is reached after the register hole is opened depends on the time at which it is triggered.

Figure~\ref{fig:gz} and Table~\ref{tab:widthMax} show that the width of this transition region, denoted by $\Delta \gamma_\mathrm{max}$, generally increases with the embouchure parameter $\zeta$, at least for $\zeta \geq 0.1$. 
However, for the tightest embouchure ($\zeta = 0.05$), this trend no longer holds. 
This deviation can be attributed to the fact that the multistable regimes differ between high and low values of $\zeta$.

As shown on Figure~\ref{fig:propR2}(a), when $\zeta = 0.05$, the transition occurs for $\gamma<1$, where the equilibrium R0 is unstable. 
As a result, multistability between R2$^{(o)}$ and R0 is impossible. 
Instead, the transition involves two oscillatory regimes: the second register R2$^{(o)}$ and the first register R1$^{(o)}$. 
R2$^{(o)}$ dominates for lower $\gamma$, while R1$^{(o)}$ takes over at higher $\gamma$. 
A direct Hopf bifurcation finally transforms R1$^{(o)}$ into R0 at $\gamma = 1$ (not shown here).

For looser embouchures ($\zeta \geq 0.1$), the situation is different. 
Since $\gamma^\mathrm{III} > 1$, both R2$^{(o)}$ and R0 are stable, enabling a multistable transition between these two regimes, as illustrated in Figure~\ref{fig:propR2}(b). 
While no case of tristability (involving R0, R1$^{(o)}$, and R2$^{(o)}$) was found, such a scenario might occur when $\gamma^\mathrm{IV}$ becomes slightly greater than 1, around $\zeta \approx 0.07$.

\subsubsection{Phase-tipping and noise-induced effects}\label{sec:results:pt}

Figure~\ref{fig:lc} shows the role of the timing of the  register hole opening in the multistable transition region. 
It shows, in the projection of the system's phase space on the plane $(\hat p_1, \dot{\hat p}_1)$, how the final regime (R0 or R2$^{(o)}$) depends on the position of the initial condition along the limit cycle of R1$^{(c)}$.
These results are shown for a relaxed embouchure ($\zeta=0.4$).

For example, at $\gamma = 1.44$, only a small section of the limit cycle (near angle $3\pi/4$) leads to the equilibrium R0, while other initial conditions still converge to R2$^{(o)}$. 
This behavior is an instance of phase-tipping \cite{alkhayuon2021phase}, where the result of a transition depends not only on the perturbation but also on the phase at which it is applied.
As $\gamma$ increases, more regions of the limit cycle start converging to R0, and the basin of attraction for R0 expands on the limit cycle.

The top-center panel of Figure \ref{fig:lc} shows that the set of initial conditions leading to R0 is not necessarily connected.
This highlights the complex structure of the basin of attraction of the equilibrium in the $(\hat p_1, \dot{ \hat p }_1)$ plane.
At $\gamma=1.46$ (top right panel), only a small portion of the limit cycle, around an angle of $-\pi/3$, leads to R2$^{(o)}$.

This visualization suggests that opening the register hole at the same time in every simulation is not ideal.
If the hole is always opened at the same phase of the limit cycle (for example, near $-\pi/3$), the model will consistently produce R2$^{(o)}$.
This remains true even when R2$^{(o)}$ is produced in minority, as shown in the top-right panel of Figure~\ref{fig:lc} for $\gamma=1.46$.

\begin{figure}[H]
	\centering
	\includegraphics[width=.9\textwidth]{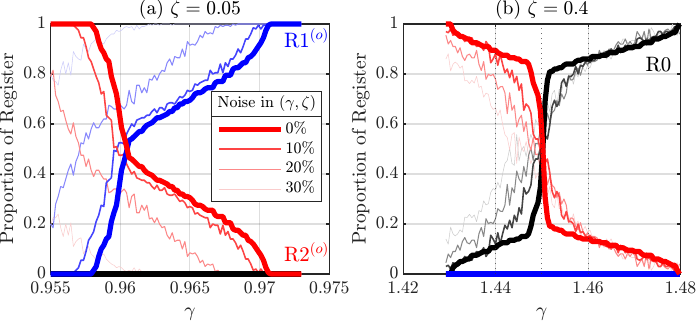}
	\caption{Evolution of the proportion of the second register (R2$^{(o)}$) obtained at the opening of the hole, for a tight embouchure $(\zeta=0.05)$ and a loose embouchure $(\zeta=0.4)$,  in the transition region $\gamma \in [\gamma^\mathrm{III}, \gamma^\mathrm{IV}]$.
	Red: second register R2$^{(o)}$. Blue: first register R1$^{(o)}$. Black: equilibrium R0.
	The lighter color shades of the curves correspond to additional noise on the control parameters.
	}
	\label{fig:propR2}
\end{figure}

\begin{figure}[H]
	\centering
	\includegraphics[height=.8\textwidth, angle=90]{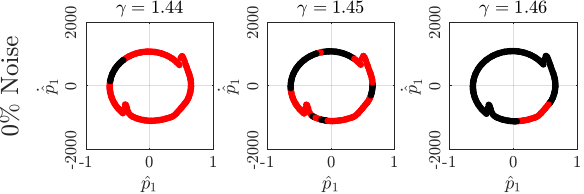}
	\includegraphics[height=.8\textwidth, angle=90]{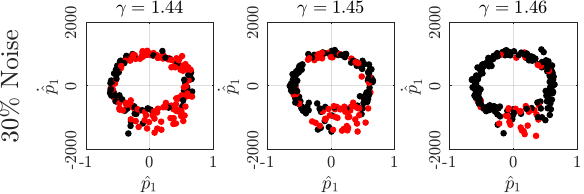}
	\caption{Positions on the limit cycle of the first register leading, when the hole is opened, to the second register (in red), and to the equilibrium (in black), for $\zeta=0.4$.
	Limit cycles are represented in the $(\hat p_1, \dot{\hat p }_1)$ space.
	Three different values of $\gamma$ are displayed on each row.
	Each row corresponds to a different amplitude of white noise added to the control parameters $(\gamma, \zeta)$ during time $t< t_\mathrm{open}$.
	First, second rows: 0~\%, 30~\% respectively.
	Other noise amplitude values are provided in Suppl.~Fig.~5.
	}
	\label{fig:lc}
\end{figure}

In Figure~\ref{fig:propR2}(b), the probability of reaching R2$^{(o)}$ follows a sigmoid curve.
Adding noise with a uniform probability density to the control parameters $(\gamma, \zeta)$ makes its central slope smoother. 
In particular, in the first half of the sigmoid, increasing noise reduces the probability of reaching R2$^{(o)}$. 
On the second half where R0 dominates, noise has little influence on the probability of reaching a given register.
This asymmetry suggests that the basins of attraction are more robust to noise in the second half of the sigmoid than in the first one.
The comparison between the two rows of Figure~\ref{fig:lc} confirms this.
For the first row, no noise is added to the control parameters.
For the second row, uniform white noise with an amplitude of 0.30\% is applied to $\gamma$ and $\zeta$.
For $\gamma = 1.44$ (first half of the sigmoid), several black dots (R0) appear all around the cycle when noise increases.
For $\gamma = 1.46$ (second half), only a few red dots (R2$^{(o)}$) are added around angle $\pi/2$ when noise increases.

Finally, for a tight embouchure ($\zeta=0.05$, Figure~\ref{fig:propR2}(a)), the evolution of the proportion of R2$^{(o)}$ with respect to $\gamma$ does not follow a symmetrical sigmoid curve  as for the relaxed embouchure ($\zeta=0.4$, Figure~\ref{fig:propR2}(b)).
We notice that when $\zeta$ decreases from 0.4 to 0.05, the second half of the curve becomes smoother, while the first half preserves this steep drop.
Additionally, noise strongly reduces the occurrence of R2$^{(o)}$ in this configuration, as illustrated by the shift of the curve towards lower blowing pressure values.
This shift points out that the basins of attraction of R2$^{(o)}$ are even less robust  to noise for low values of $\zeta$.

\begin{table}
	\centering
	\caption{Evolution of the width of the transition region $\Delta \gamma_\mathrm{max}= \gamma^\mathrm{IV}-\gamma^\mathrm{III}$ with respect to the embouchure parameter $\zeta$.}
	\label{tab:widthMax}
	\begin{tabular}{ccccccccc}
	\hline
	$\zeta$ & 0.05 & 0.1 & 0.15 & 0.20 & 0.25 & 0.30 & 0.35 & 0.40\\
		$\Delta \gamma_\mathrm{max} \cdot 10^{2}$ & 1.3 & 0.58 & 1.1 & 1.5 & 2.4 & 3.5 & 4.4 & 5.0\\ \hline
	\end{tabular}
\end{table}

\subsubsection{Long transients at low blowing pressure}\label{sec:results:transTime}

For a simulation duration of $t_\mathrm{max} = 4.0$~s, quasi-periodic regimes are mainly observed within the transition region corresponding to a low blowing pressure, defined by $\gamma \in [\gamma^\mathrm{I}, \gamma^\mathrm{II}]$. 
When $t_\mathrm{max}$ is increased, it becomes clear that these quasi-periodic regimes are in fact transient states that eventually evolve toward one of the stable regimes: R1$^{(o)}$, or R2$^{(o)}$.

\begin{figure}[H]
	\centering
	\includegraphics[width=\textwidth]{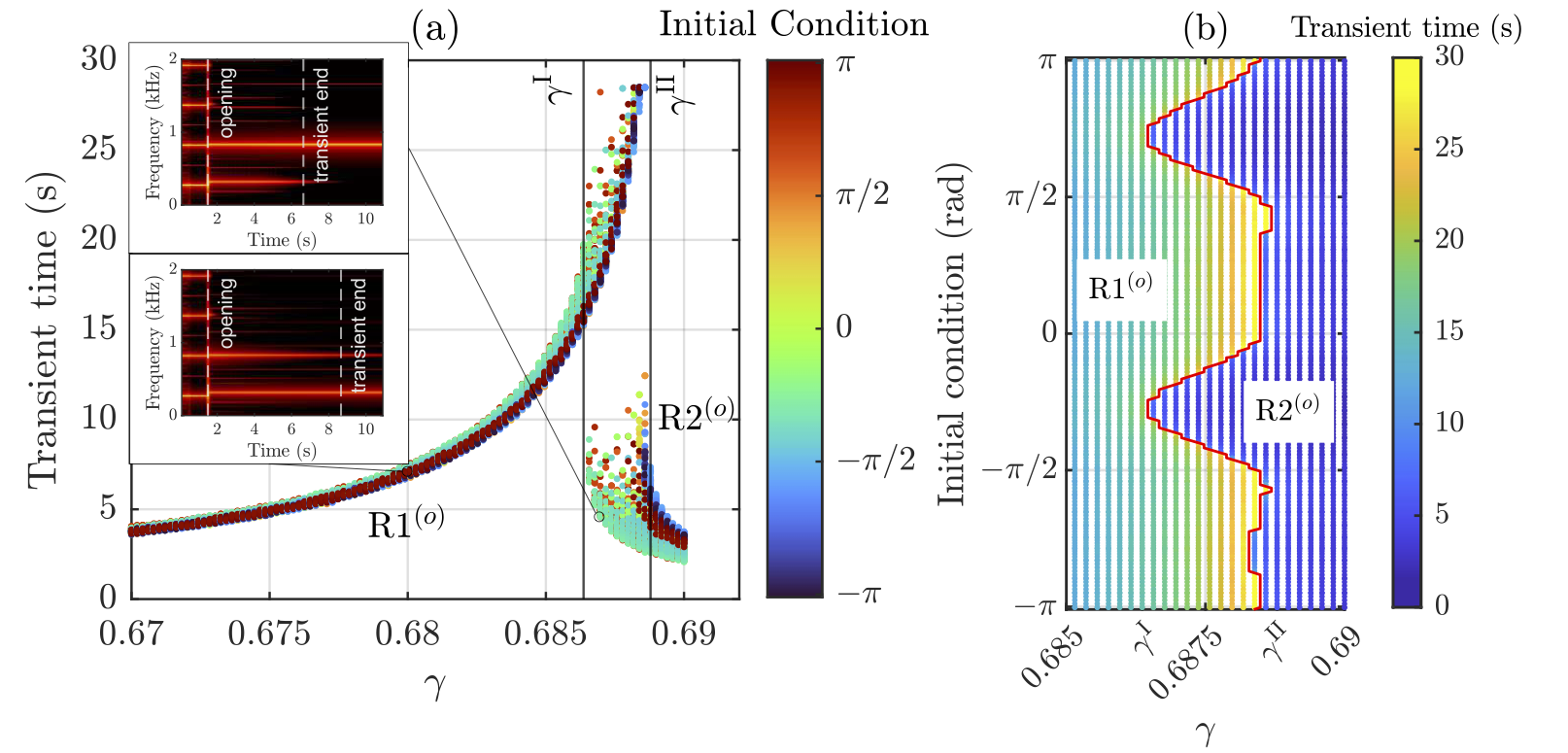}
	\caption{
	Duration between the opening of the register hole and the onset of a stable periodic regime, with respect to $\gamma$, for $\zeta=0.05$ and $\gamma <\gamma^\mathrm{II}$.
		(a): the y-axis refers to the transient time, while colors refer to the phase of the limit cycle at which the hole is opened. 
	Two additional views at the top left corner show spectrograms of $\hat p_\mathrm{in}$ leading to R1$^{(o)}$ ($\gamma=0.680$) and to R2$^{(o)}$ ($\gamma=0.687$).
	(b): the y-axis refers to the phase of the limit cycle while colors refer to the transient time. 
	The red contour separates simulations leading to R1$^{(o)}$ from simulations leading to R2$^{(o)}$.
	}
	\label{fig:transient}
\end{figure}

Figure~\ref{fig:transient} illustrates how the duration of these transient regimes increases within this region, for a fixed value of $\zeta = 0.05$. 
In the interval $\gamma \in [0.670, 0.686]$, the system consistently converges to the first register. 
However, the duration of the transient increases strongly with $\gamma$, following a hyperbolic trend. 
The average transient lasts 3.8~s at $\gamma = 0.670$ and extends to 16.4~s at $\gamma = 0.686$.

A multistable transition zone appears for $\gamma \in [0.686, 0.689]$ between R1$^{(o)}$ and R2$^{(o)}$. 
Within this range, transients leading to R1$^{(o)}$ can be extremely long, ranging from 16~s to 28~s, whereas those leading to R2$^{(o)}$ remain shorter than 12.5~s.
Beyond the upper bound of the transition region ($\gamma > \gamma^\mathrm{II}$), the duration of the transients begins to decrease.
The Supplementary Figure~6 illustrates this phenomenon experimentally, with a transient duration of 4.15~s leading to R2$^{(o)}$.

For transients ending in R1$^{(o)}$, we observe that increasing $\gamma$ also increases the standard deviation of the mean transient duration. 
Furthermore, the initial condition plays a significant role in the duration of the transient. 
The longest and shortest transients consistently come from the same angular positions on the R1$^{(c)}$ limit cycle. 
In particular, initial conditions at angles $-\pi/4$ and $3\pi/4$ produce longer transients compared to those at $-3\pi/4$ and $\pi/4$ (see also Fig.~\ref{fig:transient}(b)). 
The opposite behavior is observed for initial conditions leading to R2$^{(o)}$. Transients are shortest when starting at angles $-\pi/4$ or $3\pi/4$, and the standard deviation of the mean transient duration decreases as $\gamma$ increases.

\subsection{Discussion: on the physical phenomena underlying the experimentally observed transition regions}\label{sec:discussion}

The experiment and the simulations both show that the stability region of R2$^{(o)}$ is larger than the region where R2$^{(o)}$ is obtained from a register jump from R1$^{(c)}$.
Simulations highlight that the basin of attraction of a concurrent regime (R0 or R1$^{(o)}$) progressively encloses the limit cycle of R1$^{(c)}$.
Thus, it can be questioned whether phase-tipping would be responsible for the existence of the probabilistic regions observed in the experiment.
However, there are many differences between the experiment and the simulation.

\subsubsection{Continuous hole openings}

A hole opening is a continuous process during which the input impedance of the resonator shifts continuously from the closed hole to the open hole \cite{guillemain2005dynamic}. 
This transition may last from 20~ms \cite{taillard2018phd} to 100~ms \cite{guillemain2005dynamic}.
In the present experiment, the measured opening duration ranges between 30~ms and 50~ms (Supp.~Fig.~9).
However, a limit cycle of R1$^{(c)}$ has a period of 3.6~ms.
Consequently, a hole opening procedure lasts at least for 5.5 rotations of the R1$^{(c)}$ limit cycle.
Furthermore, controlling the hole opening time in a repeatable way would add considerable complexity to the experiment.

In the Supplementary Material (Sec.~6.1), we propose an alternative approach that consists in varying the nonlinear loss coefficient, $\hat{K}_\mathrm{nl}$, during the transition time $\Delta t^{(c) \to (o)}$. 
This method enables a smooth evolution from the closed-hole state ($\hat{K}_\mathrm{nl} \to +\infty$) to the open-hole state ($\hat{K}_\mathrm{nl} = 0.1$), without modifying the hole geometry (Suppl.~Fig.~7). 
Simulations are shown in Supplementary Figure~10 with different opening transition durations performed in the transition regions.

These results highlight a rate-induced tipping phenomenon, as studied by Terrien et al.~(2025) \cite{terrien2025basins}. 
The obtained regime does not only depend on the phase at which the hole is opened, but also on the duration of the opening process.
In particular, results suggest that slower hole openings promote register changes at higher blowing pressure. 
Similarly, for the low blowing pressure transition region $(\gamma \in [\gamma^I, \gamma^{II}]$,  Suppl.~Fig.~11), increasing the opening duration reduces the transient time leading to the second register, and shifts the transition region toward lower values of the blowing pressure.

\subsubsection{Noisy experimental blowing pressure signal}

In the experiment, for a constant blowing pressure within the transition region $P_\mathrm{blow} \in [P^\mathrm{III}, P^\mathrm{IV}]$, the standard deviation associated with noise is $\sigma_{P_\mathrm{blow}} \approx 15$~Pa.
Assuming uniform white noise (experimentally confirmed below 1000~Hz), this corresponds to an amplitude fluctuation of $\Delta P_\mathrm{noise} = 2\sqrt{3}\sigma_{P_\mathrm{blow}} \approx 52$~Pa, that is, about 4.7\% of the transition region width.
Consequently, given the complex evolution of the basins of attraction of the open-hole regimes (Fig.~\ref{fig:lc}), a phase-tipping phenomenon would be challenging to isolate.
Nevertheless, in the transition region, the unpredictability of the resulting register reflects a competition between basins of attraction in the vicinity of the R1$^{(c)}$ limit cycle.

\subsubsection{Limits of the minimal model for basin-of-attraction analysis}

Finally, the model studied in this article benefits from a minimal complexity. 
Localized nonlinear losses in the register hole are the only complex feature, essential to the production of the second register \cite{szwarcberg2024second}. 
However, this model is defined under a delayed lines formalism, which makes the study of the phase space tricky.
Constraining the set of initial conditions on the R1$^{(c)}$ limit cycle enables to bypass the explicit definition of the initial conditions in the phase space.
The authors defined a model by modal decomposition of the input impedance which accounts for localized nonlinear losses in the register hole \cite{szwarcberg2024second}.
The formalism of this model would enable to study thoroughly the evolution of the basins of attraction with respect to the control parameters.
However, this model is more resource-intensive, making such a study beyond the scope of this article.

\section{Conclusion}

An experiment is carried out on a cylindrical clarinet with a register hole.
The blowing pressure ranges for which the first and the second registers are stable are determined. 
In particular, for the open hole configuration, the multistability regions of the oscillating registers and the equilibrium are quantified.
The experiment is reproduced digitally by a waveguide model with localized nonlinear losses in the register hole.

Results indicate experimentally and numerically that within specific regions of the control parameter space, repeatedly opening the register hole from the first register can lead the system to either the second register or another regime. 
In the simulations, this probabilistic behavior is reflected in the phase space by the structure of the basins of attraction, which progressively enclose the limit cycle of the first register.
The shifting probability of convergence to a given regime directly corresponds to changes in the shape of these basins.

The robustness of the basins of attraction is studied by introducing white noise to the control parameters before opening the hole.
Results suggest that the basins are more robust to noise when multistability occurs between the equilibrium and the second register, compared to multistability between the first and the second registers.

Furthermore, long-lasting quasiperiodic regimes are investigated. 
They hide a narrow transition region in which the transient duration varies dramatically with respect to the phase at which the register hole is opened.

Finally, a model of hole opening transitions is proposed.
The results indicate that a longer opening duration favors the production of the second register and shortens the transient unstable regimes.

\section*{Declaration of competing interest}
The authors have no conflicts to disclose.

\section*{Acknowledgments}
This study has been supported by the French ANR LabCom LIAMFI (ANR-16-LCV2-007-01). 
The authors warmly thank P.\ Bindzi and V.\ Long for manufacturing the resonator, as well as J.-P.\ Dalmont for his wise experimental advice.


\end{document}